\documentclass[a4paper,UKenglish,cleveref, autoref, thm-restate]{lipics-v2021}
\category{short-papers}

\usepackage{graphics}
\title{Uncertainty Quantification in the Road-level Traffic Risk Prediction by Spatial-Temporal Zero-Inflated Negative Binomial Graph Neural Network(STZINB-GNN)} 

\titlerunning{STZINB-GNN: Uncertainty Quantification in the Road-level Traffic Risk Prediction} 

\author{Xiaowei Gao}{SpaceTimeLab, University College London (UCL), UK }{ucexwg@ucl.ac.uk}{https://orcid.org/0000-0003-3273-7499}{}

\author{James Haworth}{SpaceTimeLab, University College London (UCL), UK }{j.haworth@ucl.ac.uk}{}{}

\author{Dingyi Zhuang}{Department of Urban Studies and Planning, Massachusetts Institute of Technology (MIT), USA }{dingyi@mit.edu}{https://orcid.org/0000-0003-3208-6016}{}

\author{Huanfa Chen}{The Bartlett Centre for Advanced Spatial Analysis, University College London (UCL), UK }{huanfa.chen@ucl.ac.uk}{https://orcid.org/0000-0002-4518-7601}{}

\author{Xinke Jiang\footnote{Corresponding Author}}{School of Computer Science, Peking University (PKU), China}{thinkerjiang@foxmail.com}{}{}

\authorrunning{XW.Gao, et al.} 

\Copyright{Xiaowei Gao, et al. CC-BY} 



\relatedversion{} 

\nolinenumbers 

\EventEditors{John Q. Open and Joan R. Access}
\EventNoEds{2}
\EventLongTitle{42nd Conference on Very Important Topics (CVIT 2016)}
\EventShortTitle{CVIT 2016}
\EventAcronym{CVIT}
\EventYear{2016}
\EventDate{December 24--27, 2016}
\EventLocation{Little Whinging, United Kingdom}
\EventLogo{}
\SeriesVolume{42}
\ArticleNo{23}

\begin{document}

\maketitle

\begin{abstract}
Urban road-based risk prediction is a crucial yet challenging aspect of research in transportation safety. While most existing studies emphasize accurate prediction, they often overlook the importance of model uncertainty. In this paper, we introduce a novel Spatial-Temporal Zero-Inflated Negative Binomial Graph Neural Network (STZINB-GNN) for road-level traffic risk prediction, with a focus on uncertainty quantification. Our case study, conducted in the Lambeth borough of London, UK, demonstrates the superior performance of our approach in comparison to existing methods. Although the negative binomial distribution may not be the most suitable choice for handling real, non-binary risk levels, our work lays a solid foundation for future research exploring alternative distribution models or techniques. Ultimately, the STZINB-GNN contributes to enhanced transportation safety and data-driven decision-making in urban planning by providing a more accurate and reliable framework for road-level traffic risk prediction and uncertainty quantification. 
\end{abstract}

\section{Introduction}

In recent years, the field of traffic risk prediction has attracted considerable attention from researchers and policymakers, driven by the need to create resilient urban traffic systems and enhance their reliability in response to mounting challenges such as minimizing urban congestion, improving road safety, and making informed investments in urban infrastructure. Additionally, the zero-inflated nature of accident data, characterized by sparse incidents, poses a substantial challenge to prediction efforts.

Deep learning models have emerged as promising tools in this domain, incorporating multivariate spatiotemporal information and utilizing point-processing estimation during training to forecast traffic accidents or risk situations in subsequent time series \cite{liu2022bayesian,zhuang2022uncertainty}. For instance, de Medrano et al. \cite{de2021new} proposed a novel SpatioTemporal Neural Network (STNN) framework based on Recurrent Neural Network (RNN) to predict the number of accidents in each region of Madrid, Spain using a 5-hour prediction window. Their results show a more accurate prediction than the traditional linear statistics models as well as machine learning methods. Ren et al. \cite{ren2018deep} also employed RNN to analyze spatial and temporal patterns of traffic accident frequency and predict grid-level daily risk situations. However, the RNN model is limited by its focus on short-term temporal embedding information and its inability to fully capture the spatial heterogeneity of traffic accidents. Furthermore, Najjar et al. \cite{najjar2017combining} employed Convolutional Neural Networks (CNNs) to combine urban information from satellite imagery and open traffic accident data, mapping city-wide risk situations. Despite this, their approach neglected temporal information and faced limitations due to the quality of street image data. 

Recognizing the potential of graph neural networks (GNNs) to account for the natural connection of spatial units, researchers have proposed graph-based models for traffic risk forecasting. Zhang et al. \cite{zhang2020multi} employed a multi-modal approach to jointly consider text data from social media and imagery data from satellites, subsequently mapping grid-level traffic accident frequency using GNNs. Zhou et al. \cite{zhou2020riskoracle} introduced a novel Differential Time-varying Graph Convolution Network (GCN) to dynamically capture traffic variations and inter-subregion correlations, also predicting grid-level traffic risk. After that, they further refined their algorithms to account for hierarchical spatial dependencies, allowing for the mapping of finer grid-level urban traffic risk \cite{zhou2020foresee}. While their work addressed the zero-inflation problem of sparse accident data, their models sidestepped this challenge with a grid level and still faced limitations when predicting at the road level, which is a much finer micro-level unit compared. 

Despite the valuable foundation provided by predicted average grid-based risk levels, a significant concern in understanding traffic risk prediction is the quantification of uncertainty by considering distribution rather than mean values \cite{zhou2021stuanet,qian2022uncertainty,zhuang2022uncertainty,wu2021inductive}. Uncertainty is pervasive in urban mobility systems and plays a crucial role in accounting for potential variations in prediction results, which may arise from the aleatoric uncertainty of imbalanced risk data or the epistemic uncertainty of black-box prediction models \cite{liu2022bayesian}. Qian et al. \cite{qian2022uncertainty} explored uncertainty quantification in traffic forecasting by training a graph-based deep learning model to fit aleatoric uncertainty and combining Monte Carlo dropout with Adaptive Weight Averaging re-training methods to estimate epistemic uncertainty. Zhou et al. \cite{zhou2021stuanet} considered the irregular patterns in human mobility data as aleatoric uncertainty and the average potential variations in predicted results among specific and neighbouring regions as epistemic uncertainty. By recognizing the reducible nature of predicted epistemic uncertainty, they improved prediction performance through a gated mechanism to calibrate the predicted mobility results. Although those two approaches demonstrated the potential of combining variational inference and deep spatial-temporal embedding for predicting various distributions, they did not thoroughly investigate the sparse traffic data scenario. Zhuang et al. \cite{zhuang2022uncertainty} and Wang et al. \cite{wang2023uncertainty} highlighted the importance of considering zero-inflated distribution statistical models for analyzing sparse traffic demand data. These models offer a more accurate spatiotemporal representation of the underlying uncertainty structure suitable for incorporation with deep learning models.

Despite the progress in the field of uncertainty quantification in urban traffic research, the current focus predominantly lies on traffic demands and human mobility. This research mostly utilizes sequence or time-series data and employs coarser resolution prediction approaches, such as grid-based analysis.  This also leaves a notable research gap in terms of an investigation into finer resolution models, particularly from a road safety perspective. Road safety prediction involves non-stable and event-based characteristics, which deviate from the usual time-series data analysis. Expanding research in this area could provide a more nuanced understanding of traffic risk prediction and its significant potential for improving urban transportation safety and resilience. Filling this research gap will contribute to the holistic development of urban transportation studies, enhancing not only predictive accuracy but also the applicability of results in practical, real-world scenarios.

Building upon the work of Zhuang et al. \cite{zhuang2022uncertainty}, this paper presents the Spatial-Temporal Zero-Inflated Negative Binomial Graph Neural Network (STZINB-GNN) model, specifically designed to tackle the existing limitations in traffic risk prediction and uncertainty quantification. 

\begin{alphaenumerate}
    \item The zero-inflated negative binomial model is employed to effectively distinguish between non-risk and risk levels across road segments.
     \item The spatial-temporal Graph Neural Network (ST-GNN) is responsible for learning and fitting the parameters of probabilistic distributions. 
\end{alphaenumerate}

To the best of our knowledge, this is the first attempt to merge these two elements for road-level risk estimation. Empirical evidence showcases the enhanced performance of our proposed model when applied to road-daily resolution traffic risk data. 

This paper is structured as follows: Section 2 describes the development of the model and provides detailed explanations of its components. Section 3 presents the dataset employed for the case study, outlines the evaluation metrics, and discusses the experimental results. Finally, Section 4 offers conclusions and highlights potential avenues for future research.

\section{Methdology}

\begin{figure*}[htbp]
    \centering
    \includegraphics[width=1\linewidth]{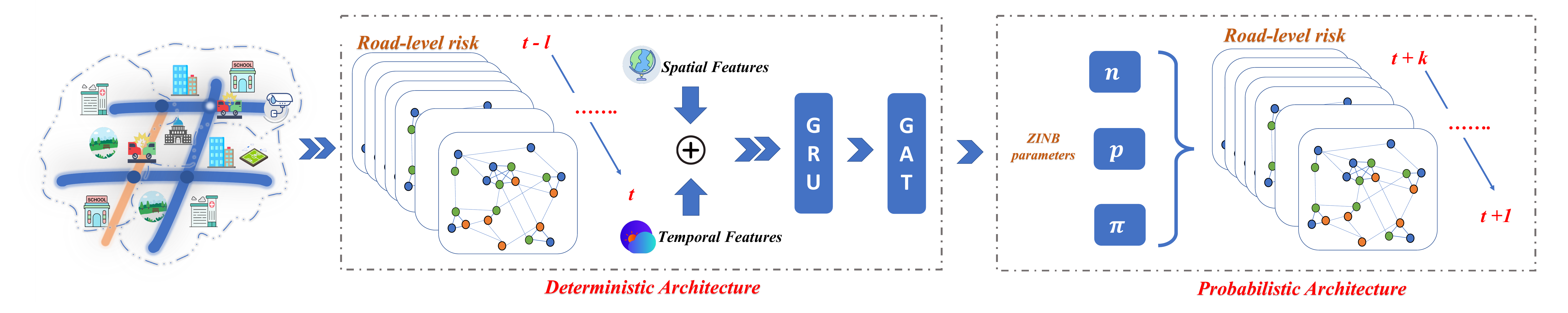}
    \caption{Framework of STZINB-GNN model.}
    \label{fig:model}
\end{figure*}

Our objective is to predict future traffic risk and associated confidence intervals for each individual road segment across $k$ forthcoming time intervals, using $m$ roads' risk conditions from previous time windows lasting $T$ days. This is a sequence-to-sequence prediction problem, as illustrated in Figure \ref{fig:model}. We construct the road graph, $\mathcal{G} = (V, E)$, where $V$ represents the set of roads, and $E$ denotes the edges connecting these roads. The adjacency matrix $A\in \mathbb{R}^{V\times V }$ indicates the relationships between roads, and $|V|=m\times m$. More specifically, $x_{it}$ signifies the risk level of the $i^{th}$ road segment during the $t^{th}$ time interval, where $i\in V$, $x_{it}\in \mathbb{N}$. Subsequently, $X_{t}\in \mathbb{N}^{|V|\times T}$ designates the risk conditions for all road segments within the $t^{th}$ time interval, with $x_{it}$ as its component. Our aim is to employ historical records $X_{1:t}$ and spatial-temporal features as input data to estimate the distribution of predicted $X_{t:t+k}$ (i.e., the risk levels for each road over the next $k$ time intervals), thus examining the anticipated value and confidence intervals of the future risk situation.

First, we assume that the inputs follow the ZINB distribution with probability mass function as:
\begin{equation}
    \label{eq:zinb}
    f_{ZINB}(x_k;\pi,n,p) = \left\{ \begin{array}{cc}
        \pi + (1-\pi)\left( \begin{array}{c}
        x_k+n-1   \\
        n-1
    \end{array} \right) (1-p)^{x_k}p^n & \text{if } x_k=0  \\
        (1-\pi)\left( \begin{array}{c}
        x_k+n-1   \\
        n-1
    \end{array} \right) (1-p)^{x_k}p^n & \text{if } x_k>0
    \end{array}\right.,
\end{equation}
where $\pi$ is the inflation of zeros, $n$ and $p$ determine the number of successes and the probability of a single failure respectively. $x_k$ denotes a road's traffic risk level at one time. All three parameters $\pi,n,p$ are learned by spatial-temporal GNNs (STGNN), where the temporal encoder applies a Gated Recurrent Unit (GRU) and then the spatial encoder applies Graph Attention Network (GAT):
\begin{equation}
    n_{t+1:t+k}, p_{t+1:t+k}, \pi_{t+1:t+k} = STGNN(X_{1:t}; F_{1:t}; A) = GAT\bigl(GRU(X_{1:t}; F_{1:t}); A \bigl).
\end{equation} 

Here, $F_{t}$ represents the spatiotemporal features of roads on the $t^{th}$ day, while $X_{t}$ corresponds to the risk level observed on the same day. This relationship illustrates the connection between road features and risk levels at specific roads in time.

The log-likelihood of ZINB is composed of the  $y=0$ and $y>0$ parts, and can be approximated as follows:

\begin{equation}
 LL_y= 
\left\{
\begin{array}{lccc}
\log \pi+\log (1-\pi) p^n & \text{when} & y=0 & \\[6pt]
\log (1-\pi) +\log \Gamma(n+y)-\log \Gamma(y+1) & & & \\
-\log \Gamma(n)+n \log p+y \log (1-\pi) & \text{when} & y>0 &
\end{array}
\right.,
\end{equation}

where $y$ is the ground-truth value, $\Gamma$ is the Gamma function and $n, p, \pi$ is learned by STGNN. The final negative log-likelihood loss function is given by:
\begin{equation}
    NLL_{STZINB} = - LL_{y=0} - LL_{y>0}.
\end{equation}

\section{Result}
We evaluated the model's performance using a real-world dataset from Lambeth Borough in London, UK. This dataset comprises 5,659 road segments and a total of 1,335 accidents throughout 2019\footnote{https://roadtraffic.dft.gov.uk/downloads}. We calculated a daily risk level by combining the number of accidents with the severity of each accident. We then identified the nearest road segment and accounted for the spillover effects on first and second-order neighbouring roads \cite{zhou2020riskoracle} to assign each road segment a risk value for each day. Notably, the zero-inflation rate for road-level accident risk in Lambeth Borough is 98.7\%, indicating that a significant proportion of the road segments experienced no accidents during the observed period.

The evaluation metrics for assessing the model performance are categorized into four aspects. Traditional accuracy measures, including Mean Absolute Error (MAE), Mean Absolute Percentage Error (MAPE), and Root Mean Squared Error (RMSE), evaluate the model's ability to accurately predict risk values. Uncertainty quantification is assessed using the Kullback-Leibler Divergence (KLD), which indicates how closely the distribution of the model's output risk values approximates the distribution of the true risk values. Lower values for these above metrics are desirable, as they signify a smaller difference between the predicted and actual risk values as well as distributions. 

The true-zero rate (ZR) quantifies the model's capacity to accurately replicate the sparsity observed in the ground-truth data. Additionally, the Hit Rate (HR) is assessed based on information entropy. To compute HR, we first select the top 20\% of road segments with the highest predicted risk values and then consider the predicted risk values' information entropy, which is derived from the KLD uncertainty quantification. We calculate the HR by selecting those road segments with lower predicted uncertainty among the top 20\% risk roads, where the road information entropy is below the mean value of the entire road's entropy.

Higher ZR and HR suggest better model performance in identifying road segments with no accidents and those with accidents, respectively.

\begin{table}[ htbp]
\footnotesize
\resizebox{\columnwidth}{!}{
\begin{tabular}{cc|cccccccc}
\hline
\multicolumn{2}{c|}{Results}                                & \multicolumn{4}{c|}{Long(14-14)}                       & \multicolumn{4}{c}{Short(7-7)}    
                                                              \\ \hline
\multicolumn{1}{c|}{Metrics}                   & Models     & STZINB & STG  & STNB   & \multicolumn{1}{c|}{HA}     & STZINB & STG  & STNB   & HA     \\ \hline
\multicolumn{1}{c|}{\multirow{3}{*}{ACC}}      & MAE        & \textbf{0.054} & {0.118} & \underline{0.080} & \multicolumn{1}{c|}{0.135} & \textbf{0.077} & \underline{0.048} & 0.105 & 0.134 \\
\multicolumn{1}{c|}{}                          & MAPE       & \textbf{0.026} & 0.405 & \underline{0.036} & \multicolumn{1}{c|}{0.414} & \textbf{0.025} & 0.443 & \underline{0.078} & 0.485 \\
\multicolumn{1}{c|}{}                          & RMSE       & \textbf{0.119} & 0.183 & \underline{0.139}  & \multicolumn{1}{c|}{0.211} & \textbf{0.139} & \underline{0.185} & 0.172 & 0.238 \\ \cline{1-2}
\multicolumn{1}{c|}{Uncertainty}               & KLD        & \textbf{0.259} & {0.504} & 1.558  & \multicolumn{1}{c|}{\underline{0.269}} & \underline{0.473} & {0.522} & 0.759 & \textbf{0.264} \\ \cline{1-2}
\multicolumn{1}{c|}{Zero Inflated}             & ZR         & \textbf{0.641} & 0.199 & \underline{0.562} & \multicolumn{1}{c|}{0.520} & \textbf{0.579} & 0.102  & \underline{0.518} & 0.503 \\ \cline{1-2}
\multicolumn{1}{c|}{\multirow{1}{*}{Hit Rate}} & HR\@20\% & \textbf{0.618} & {0.267} & \underline{0.447} & \multicolumn{1}{c|}{0.452}  & \textbf{0.575} & {0.301} & 0.442 & \underline{0.443}   \\ \hline
\end{tabular}
}
\caption{Model Results for the Lambeth Borough, London}
\end{table}

In the table, bold fonts mean the best values among all the baseline models while the underline means the second-best values. The baseline models used for comparison include Spatial-temporal Graph Neural Network with Gaussian Distribution (STG), Spatial-temporal Graph Neural Network with Negative Binomial Distribution (STNB), and Historical Average (HA). It is evident that our proposed model outperforms the baseline models across all evaluation metrics, with the exception of KLD in short-term prediction scenarios, where it ranks second. This demonstrates the effectiveness of our model in capturing the skewed data distribution through its zero-inflated components, which leads to more accurate results and improved reliability when approximating the true risk distribution. Notice that both ZR and HR20\% metrics, which are important to measure the occurrence of events in practice, have received significant accuracy improvement. This is due to introducing the parameter $\pi$ in Equation \ref{eq:zinb}, which can effectively learn the sparsity of the data.

Furthermore, our model's capability to reliably predict higher risk values enables us to achieve an accuracy of approximately 61.8\% or 57.5\% in identifying the exact locations of accidents, which also significantly outperforms the other GNN counterparts. This highlights the potential of our approach to significantly enhance transportation safety and facilitate data-driven decision-making in urban planning.

\section{Discussion and Conclusion}

In this study, we developed a versatile spatial-temporal Graph Neural Network (GNN) framework for predicting the probabilistic distribution of sparse road traffic risk and quantifying its associated uncertainty. By employing Gated Recurrent Units (GRUs) to capture temporal correlations and Graph Attention Networks (GATs) to model spatial dependencies, we created a comprehensive framework that embeds the spatial and temporal representations of distribution parameters. These parameters are then fused to obtain a distribution for each spatial-temporal data point.

Our case study, based on the urban risk situation of Lambeth borough in the UK, demonstrated that the proposed model consistently delivers more accurate and reliable results compared to other methods. Despite its performance, the model also has certain limitations. When addressing real risk levels that extend beyond binary values, the negative binomial distribution may not be the most suitable choice. Future work could explore alternative distribution models or techniques that better capture the complexity and nuances of real-world risk levels. This would further enhance the model's applicability and predictive capabilities, ultimately contributing to improved transportation safety and data-driven decision-making in urban planning.


\bibliography{lipics-v2021-sample-article}

\end{document}